\documentstyle[aps,prd,preprint,floats,epsfig]{revtex}

\begin{document}
\draft
\tighten

\preprint{\tighten\vbox{\hbox{\bf CLNS 00-1709}
                        \hbox{\bf CLEO 00-28}}}

\title{\Large \boldmath Correlated $\Lambda_c^+{\overline\Lambda_c}^-$ 
production
in $e^+e^-$ annihilations at $\sqrt{s}\sim$10.5 GeV}

\author{CLEO Collaboration}
\date{\today}

\maketitle
\tighten

\begin{abstract}
Using 13.6 ${\rm fb}^{-1}$ of continuum two-jet
$e^+e^-\to c{\overline c}$ events
collected with the CLEO detector, we have searched for
baryon number correlations at the primary quark level.
We have measured the likelihood for a $\Lambda_c^+$ charmed baryon
to be produced in the hemisphere
opposite a ${\overline\Lambda_c}^-$
relative to the likelihood for a $\Lambda_c^+$ charmed baryon to be 
produced opposite an anticharmed meson ${\overline D}$; in
all cases, the reconstructed hadrons must have momentum greater than 
2.3 GeV/c.
We find that, given a ${\overline\Lambda_c}^-$ 
(reconstructed in five different decay modes),
a $\Lambda_c^+$ is observed in the opposite hemisphere 
$(0.72\pm0.11)$\% of the time (not corrected for efficiency).
By contrast, given a ${\overline D}$ in one hemisphere, 
a $\Lambda_c^+$ is observed in the opposite hemisphere only 
$(0.21\pm0.02)$\% of the
time. Normalized to the total number of either ${\overline\Lambda_c}^-$ or
${\overline D}$ ``tags'',
it is therefore 3.52$\pm$0.45$\pm$0.42 
times more likely to find a $\Lambda_c^+$ opposite
a ${\overline\Lambda_c}^-$ than a ${\overline D}$ meson.
This enhancement is not observed in the JETSET 7.3 $e^+e^-\to c{\overline c}$
Monte Carlo simulation.
\end{abstract}

\setcounter{footnote}{0}

\newpage

\begin{center}
A.~Bornheim,$^{1}$ E.~Lipeles,$^{1}$ S.~P.~Pappas,$^{1}$
M.~Schmidtler,$^{1}$ A.~Shapiro,$^{1}$ W.~M.~Sun,$^{1}$
A.~J.~Weinstein,$^{1}$
D.~E.~Jaffe,$^{2}$ R.~Mahapatra,$^{2}$ G.~Masek,$^{2}$
% ...
\end{center}
 
\small
\begin{center}
$^{1}${California Institute of Technology, Pasadena, California 91125}\\
$^{2}${University of California, San Diego, La Jolla, California 92093}\\
% ...
\end{center}
 

\section{Introduction}
At high energy, quark fragmentation can be calculated using Perturbative QCD
(PQCD). The results are often referred to as QCD showers
\cite{part-shower}. However, these calculations require, for comparison with
experimental results, some model or experimental input to take into account
the eventual hadronization of gluons and $q\overline q$ pairs into actual
hadrons of known mass.

 Several models have thus been developed, mostly based on a chromodynamic
string connecting the initial $q_0$ and $\overline q_0$
\cite{lund,a-m,russian,ic,nos-prd}, where
the subscript indicates that these are the primary quarks formed
by $e^+e^-$ annihilation.  When stretched because of the initial
momentum of $q_0$ and $\overline q_0$, the string breaks by creation of
secondary $q\overline q$ pairs, generating two substrings.  The process
iterates then for the new substrings until the physical hadrons are produced.
This last stage is the critical one where QCD loses its predictive power.

  The model most frequently used to simulate the process $e^+e^-\to
q_0\overline q_0\to\ hadrons$ is the QCD inspired Lund String Model
(LSM)~\cite{lund}, implemented in the Jetset Monte Carlo simulation
package~\cite{pythia}.  
In order to solve the problem of generating hadrons of known mass, 
the Jetset implementation of LSM, at
each step of its iterative process, splits a $q\overline q$ string into a
hadron of known quantum numbers and mass and a remainder string having the
leftover quantum numbers and energy-momentum.  When, finally, the remainder
string has a mass below a suitable limit, this string is made into two known
hadrons that, together, carry the leftover quantum numbers. 

In this paper we report on correlations between charm baryons
in the fragmentation of a $c\overline c$ system produced in the reaction
$e^+e^-\to c\overline c$ at $\sqrt{s}\sim10$~GeV, where the annihilation
occurs in a largely low-$Q^2$, non-perturbative regime. 
In contrast to inclusive single-particle production,
compensation of baryon and charm number is a 
more subtle aspect of quark
fragmentation modeling.  One obvious question is whether baryon compensation
occurs locally (e.g., small rapidity difference between baryon and
antibaryon) or globally (large rapidity difference).  Previous studies of
$\Lambda{\overline\Lambda}$ production at $\sqrt{s}$=90 GeV found that in
events containing both a $\Lambda$ and a ${\overline\Lambda}$, the two
particles tended to be produced at very similar rapidities~\cite{OPAL-paper}.

Consider the case in which a charm baryon is produced in the first step of
fragmentation (e.g. $e^+e^-\to c \overline c ;\ c\to\Lambda_c X$); it is
possible that both baryon and quantum charm numbers be compensated in the
opposite hemisphere (e.g. $e^+e^-\to c\overline c;\ c\to\Lambda_c X,\
\overline c\to{\overline\Lambda_c}^- X$).  In the limit that the $c$ and
$\overline c$ quarks fragment independently 
(presumably true at some sufficiently high energy),
this type of correlation is not expected.  The Jetset
implementation of the Lund model, because of the mechanism outlined above,
does not produce such a correlation. 

In the very reasonable approximation of neglecting string splitting by
$c\overline c$ tunneling from 
the vacuum, at $\sqrt{s}\sim10$~GeV all observed
charmed particles must contain a primary quark.  We take advantage of this to
discriminate between independent vs. correlated fragmentation models.  
If we assume that primary quarks fragment independently, 
then the number of times that we find 
a ${\Lambda}^{+}_c$ baryon opposite
a ${\overline\Lambda_c}^-$ 
antibaryon in an event
(i.e., $cos\theta(\Lambda_c^+,{\overline\Lambda_c}^-)<$0,
denoted ``$\Lambda_c^+|{\overline\Lambda_c}^-$''), 
scaled to the total number of observed
${\overline\Lambda_c}^-$ 
(denoted 
``$\frac{{\Lambda}^{+}_{c} | 
{\overline\Lambda_c}^{-}}{{\overline\Lambda_c}^{-}}$''), 
should be equal to the number of times that we find a 
${\Lambda_c^+}$ baryon opposite any other anti-charmed 
hadron ${\overline H_c}$,
scaled to the total number of observed anti-charmed hadrons
($\frac{{\Lambda_c^+} | \overline{H_{c}}}{\overline{H_{c}}}$). 
In this analysis, we will check the equality of these ratios; 
by comparing ratios in this way, we cancel many experimental
systematics.
Schematically, we are comparing the following event topologies:\\
\begin{center}{
\[
   \begin{array}{cccccccc}
      & & & \overline{c} &  c & & & \\
      & & {\overline\Lambda_c}^{-} \hookleftarrow & & & \hookrightarrow
          {\Lambda}^{+}_{c} & & 
   \end{array}
\]
\large
\[
   \begin{array}{cccccccc}
      & & & \it {vs} & & & 
   \end{array}
\]
\normalsize
\[
   \begin{array}{cccccccc}
      & & & \overline{c} &  c & & & \\
      & & \it{\overline{H_{c}}} \hookleftarrow & & & \hookrightarrow
          {\Lambda}^{+}_{c} & & 
   \end{array}
\]
}\end{center}

Specifically, we compare the rate of 
${\Lambda}^{+}_c |{\overline\Lambda_c}^{-}$ production to the 
rate of ${\Lambda}^{+}_c |\overline{D^0}$  and 
${\Lambda}^{+}_c |D^-$ production.  We reconstruct 
${\overline D^0}$'s and $D^-$'s through the well-measured 
decay modes ${\overline D}^0\to K^+\pi^-$ 
and $D^-\to K^+\pi^-\pi^-$, respectively.  
$\Lambda_c^+$'s are fully
reconstructed in the decay modes
$\Lambda_c^+\to pK^-\pi^+$, $\Lambda_c^+\to pK^0_s$, 
$\Lambda_c^+\to\Lambda\pi^+$, $\Lambda_c^+\to\Lambda\pi^+\pi^-\pi^+$,
$\Lambda_c^+\to pK^0_s\pi^+\pi^-$,\footnote{Charge conjugation is implicit.}
and partially reconstructed
through $\Lambda_c^+\to\Lambda$X.\footnote{According to
JETSET 7.3 simulations,
95\% of the $\Lambda$'s opposite 
an anti-charm tag are
charmed baryon daughters (predominantly $\Lambda_c$) after imposing a
1 GeV/c minimum momentum requirement
on the $\Lambda$ ($p_\Lambda>$1 GeV/c).} 
$\Sigma_c$'s are also studied; these are reconstructed in 
$\Sigma_c^{++}\to\Lambda_c^+\pi^+$ and $\Sigma_c^0\to\Lambda_c^+\pi^-$.

Under the assumption of independent fragmentation, 
we expect the relative
production ratios to satisfy:

\begin{center}
$\frac{\Lambda^+_c | {\overline\Lambda_c}^-}{{\overline\Lambda_c}^-} 
\div \frac{\Lambda^+_c | \overline{D^0}}{\overline{D^0}}$ = 1 \\
\end{center}

and,

\begin{center}
$\frac{\Lambda^+_c | {\overline\Lambda_c}^-}{{\overline\Lambda_c}^-} 
\div \frac{\Lambda^+_c | D^-}{D^-}$ = 1.
\end{center}

\section{Apparatus and Event Selection}
\label{sec:event_selection}

This analysis was performed using the CLEO II and the upgraded
CLEO II.V detectors operating at the
Cornell Electron Storage Ring (CESR) at center-of-mass energies $\sqrt{s}$
= 10.52--10.58 GeV.  
For 4.5 fb$^{-1}$ of the data used for this analysis 
(``CLEO-II'' data\cite{kubota92}),
measurements of charged particle momenta were made with
three nested coaxial drift chambers consisting of 6, 10, and 51 layers,
respectively. In a subsequent upgrade (``CLEO-II.V''\cite{CLEO-IIV}), 
the inner tracking chamber was replaced with a high-precision
silicon detector (corresponding to the remaining 9.1 fb$^{-1}$ of the data
used for this analysis).
The entire tracking system fills the volume from $r$=3 cm to $r$=1 m, with
$r$ the radial coordinate relative to the beam (${\hat z}$) axis. 
This system is very efficient ($\epsilon\ge$98\%) 
for detecting tracks that have transverse momenta ($p_T$)
relative to the
beam axis greater than 200 MeV/c, and that are contained within the good
fiducial volume of the drift chamber ($|\cos\theta|<$0.94, with $\theta$
defined as the polar angle relative to the beam axis). 
This system achieves a momentum resolution of $(\delta p/p)^2 =
(0.0015p)^2 + (0.005)^2$ ($p$ is the momentum, measured in GeV/c). 
Pulse-height measurements in the main drift chamber provide specific
ionization resolution
of 5.0\% (CLEO II.V) or 5.5\% (CLEO II) 
for Bhabha events, giving good $K/\pi$ separation for tracks with
momenta up to 700 MeV/c and separation of order 2$\sigma$ in the relativistic
rise region above 2.5 GeV/c. 
Outside the central tracking chambers are plastic
scintillation counters, which are used as a fast element in the trigger system
and also provide particle identification information from 
time-of-flight measurements.  

Beyond the time-of-flight system is the electromagnetic calorimeter,
consisting of 7800 thallium-doped CsI crystals.  The central ``barrel'' region
of the calorimeter covers about 75\% of the solid angle and has an energy
resolution which is empirically found to follow:
\begin{equation}
\frac{ \sigma_{\rm E}}{E}(\%) = \frac{0.35}{E^{0.75}} + 1.9 - 0.1E;
                                \label{eq:resolution1}
\end{equation}
$E$ is the shower energy in GeV. This parameterization includes
noise effects, and translates to an
energy resolution of about 4\% at 100 MeV and 1.2\% at 5 GeV. Two end-cap
regions of the crystal calorimeter extend solid angle coverage to about 95\%
of $4\pi$, although energy resolution is not as good as that of the
barrel region. 
The tracking system, time-of-flight counters, and calorimeter
are all contained 
within a superconducting coil operated at 1.5 Tesla. 
An iron flux return interspersed with
proportional tubes
used for muon detection is located immediately outside the coil and 
in the two end-cap regions.

The event sample 
used for this measurement is comprised of 9.1 ${\rm fb}^{-1}$ of data
collected at the $\Upsilon$(4S) resonance and 4.5 ${\rm fb}^{-1}$ of data 
collected about 60 MeV below the $\Upsilon$(4S) resonance. Approximately
$18\times 10^6$ continuum c\=c events are included in this sample.
Charged track candidates for protons, kaons or pions 
must pass the following restrictions:

(a) The track must have an impact parameter relative to the 
estimated event vertex
less than 5 mm in a plane perpendicular to the
beam axis (${\hat r}-{\hat\phi}$) and no more than 5 cm along
the beam axis. The estimated event vertex is obtained by 
averaging the $e^+e^-$ interaction point over a full run.

(b) The track has specific ionization information 
consistent (at the 99\% confidence level) with its assumed
particle identity. 

(c) The track must have momentum greater than 100 MeV/c.

All reconstructed charmed hadrons must have momentum greater than
2.3 GeV/c to ensure that there is no contamination from $B$-meson decays
to charm.

\subsection{Production Ratios}

We define the single tag yield to be the 
number of reconstructed events containing one 
anti-charmed hadron ${\overline H_c}$.
This yield (Table \ref{tab:Single-tags})
is determined by
fitting a double-Gaussian signal over a smooth, 
low-order polynomial background
functional. The signal function is the sum of two Gaussian
functions, one narrow and one broad, and is a better representation of
the expected signal line shape than a single Gaussian function.

\begin{table}[htpb]
\caption{\label{tab:Single-tags}
Single Tag Yields obtained from one-dimensional fits
(charge conjugate modes are implied);
statistical errors only are shown.}
\begin{center}
\begin{small}
\begin{tabular}{ccc}
 \quad & \quad Data & 
Monte Carlo  \\  \hline\hline
 \quad & \quad ($\sim$60 M hadronic events) & 
($\sim$160 M hadronic events) \\  \hline\hline
\quad $\Lambda_c$ (5 modes) & 70199 $\pm$ 1604 & \\
\quad $\Lambda_c$ ($pK^-\pi^+$ + $pK^0_s$ only) & $56110\pm1776$ 
& 130380 $\pm$ 1943 \\
\quad $\Sigma_c^0\to\Lambda_c^+\pi^-$ + $\Sigma_c^{++}\to\Lambda_c^+\pi^+$ 
(sum) & 3804 $\pm$ 185 & 6522 $\pm$ 182 \\
 \quad $\Lambda \to$ p$\pi^-$    & 735343 $\pm$ 1198 & 2136997 $\pm$ 1674 \\
 \quad $D^0 \to$ $K^-\pi^+$        & 352294 $\pm$ 1668 & 1045776 $\pm$ 2294 \\
 \quad $D^+ \to$ $K^-\pi^+\pi^+$   & 273597 $\pm$ 2148 & 705357 $\pm$ 3232 \\
                                                                        \hline
\end{tabular}
\end{small}
\end{center}
\end{table}

\begin{figure}[htpb]
\begin{picture}(200,250)
\includegraphics{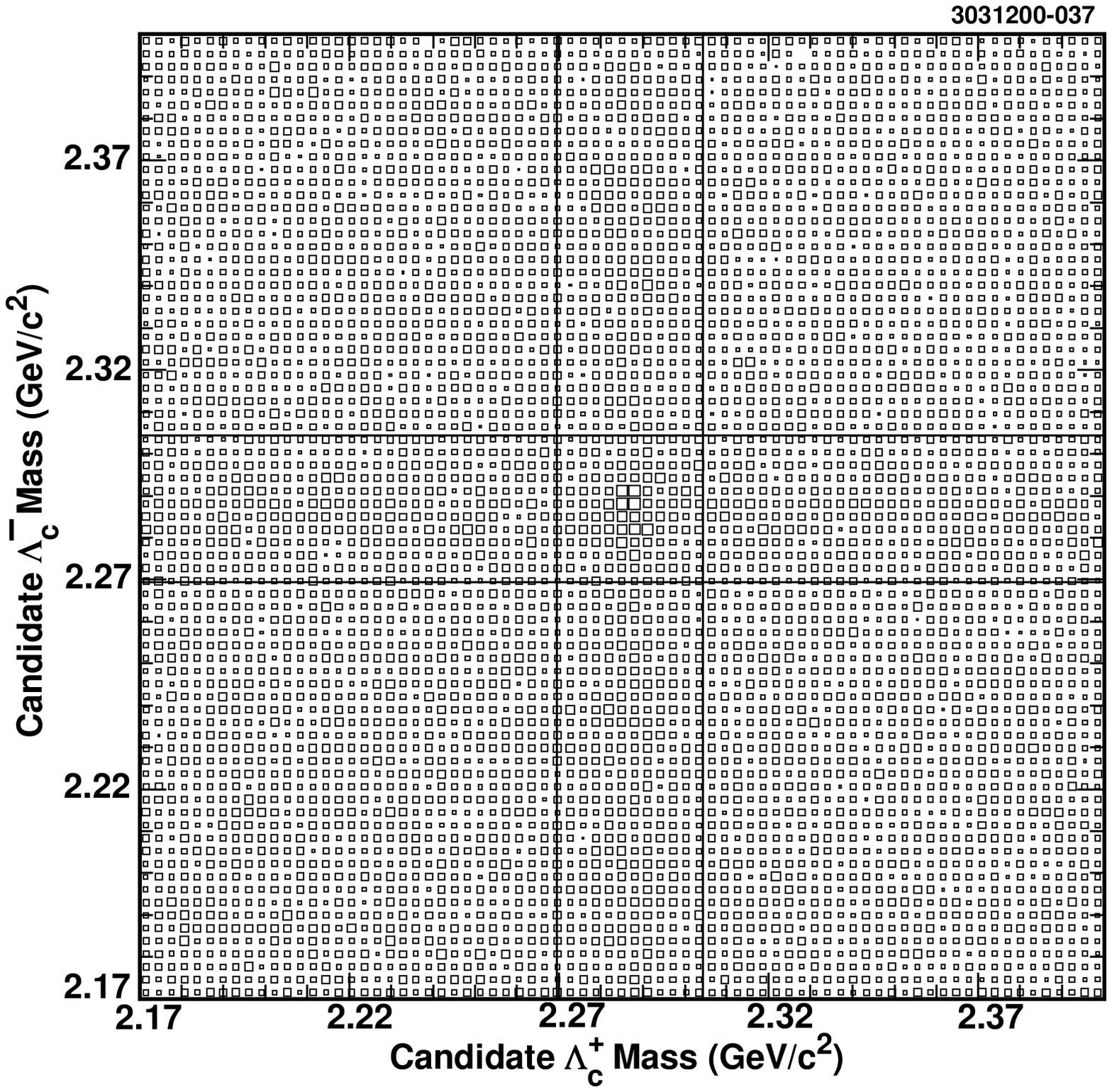}
\end{picture} \\
\caption{\small 
Double tag invariant mass
plot of $\Lambda_c^+$ candidates
plotted vs. invariant mass of ${\overline\Lambda_c}^-$ candidates 
(${\Lambda}^{+}_{c} | {\overline\Lambda_c}^{-}$) from data. Shown is the
sum of the modes: 
$\Lambda_c^+\to pK^-\pi^+$, $\Lambda_c^+\to pK^0_s$, 
$\Lambda_c^+\to\Lambda\pi^+$, $\Lambda_c^+\to\Lambda\pi^+\pi^-\pi^+$,
and $\Lambda_c^+\to pK^0_s\pi^+\pi^-$ (and their charge conjugates,
in the case of ${\overline\Lambda_c}^-$ reconstruction).
Horizontal and vertical solid lines designate signal bands.
}
\label{fig:lclcb}
\end{figure}

\begin{figure}[htpb]
\begin{picture}(200,250)
\includegraphics{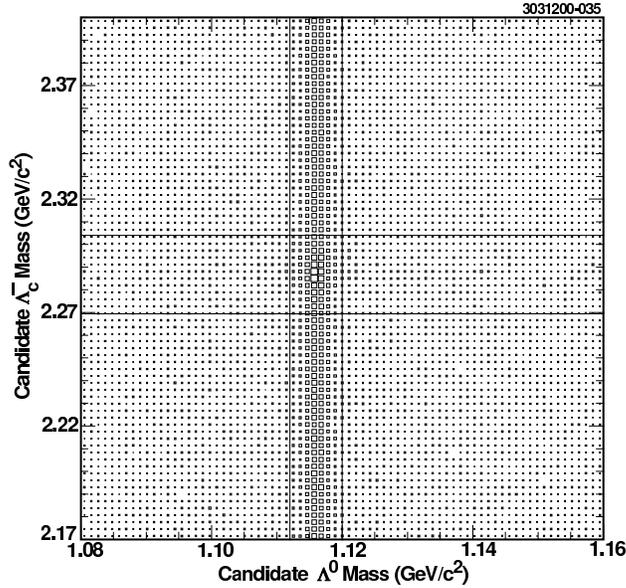}
\end{picture} \\
\caption{\small Double tag plot of 
$\Lambda | {\overline\Lambda_c}^-$ (plus charge conjugate) from data.  
The ${\overline\Lambda_c}^-$ is selected as in the previous figure; the
$\Lambda$ is reconstructed in $\Lambda\to p\pi$. 
Horizontal and vertical solid lines designate signal bands.
}
\label{fig:lcl0b}
\end{figure}

\begin{figure}[htpb]
\begin{picture}(200,250)
\includegraphics{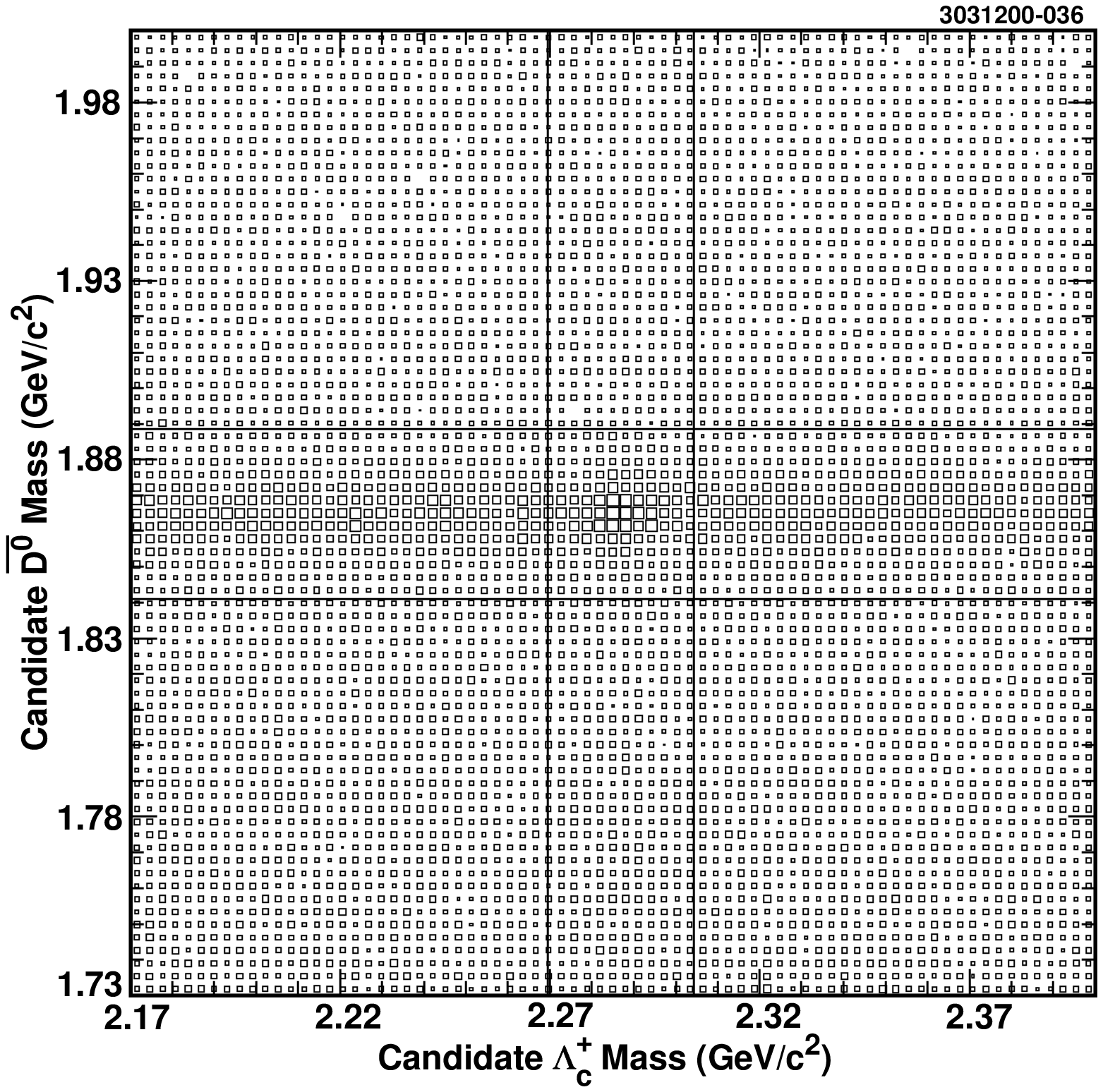}
\end{picture} \\
\caption{\small Double tag plot of $\Lambda^+_c | \overline{D^0}$
(plus charge conjugate) for data.
Horizontal and vertical solid lines designate signal bands.
}
\label{fig:lcd0b}
\end{figure}

\begin{figure}[htpb]
\begin{picture}(200,250)
\includegraphics{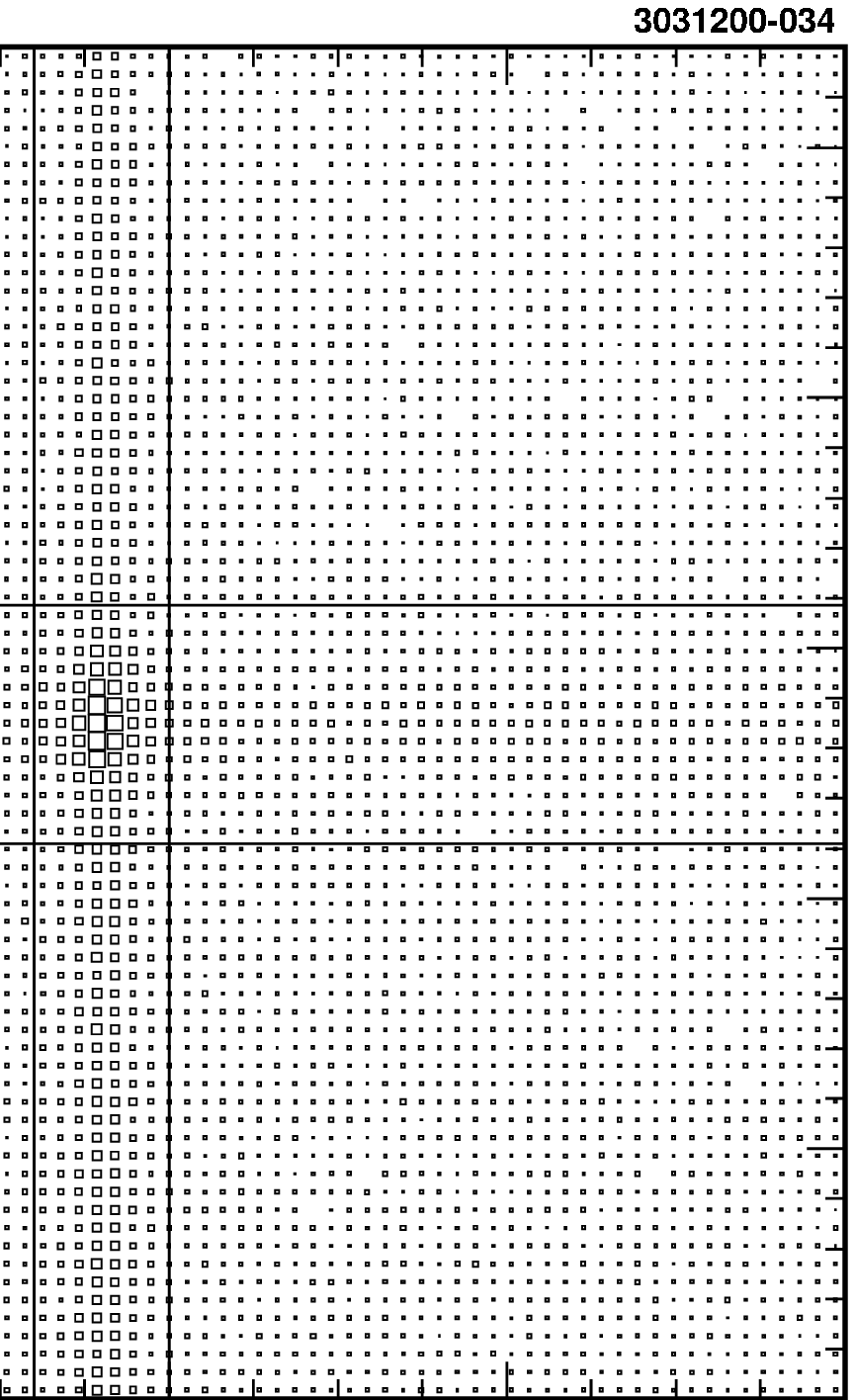}
\end{picture} \\
\caption{\small Double tag plot of 
${\Lambda} | \overline{D^0}$ (plus charge conjugate) for data.
Horizontal and vertical solid lines designate signal bands.
}
\label{fig:lamd0b}
\end{figure}

The number of double tags (Table \ref{tab:Double-tags}) is 
defined as the number of 
events in which two specific particles are reconstructed in opposite
hemispheres
(i.e., the opening angle between the two 
particles must be greater than 90 degrees).
The total
number of double tags
for $\Lambda_c^+|{\overline\Lambda_c}^-$, $\Lambda_c^+|{\overline\Lambda}$,
$\Lambda_c^+|{\overline D^0}$ and 
$\Lambda|{\overline D^0}$
is extracted from two-dimensional 
invariant mass plots, shown in Figures 
\ref{fig:lclcb}, \ref{fig:lcl0b}, \ref{fig:lcd0b}, and \ref{fig:lamd0b}
respectively.

\begin{figure}[htpb]
\begin{picture}(200,250)
\includegraphics{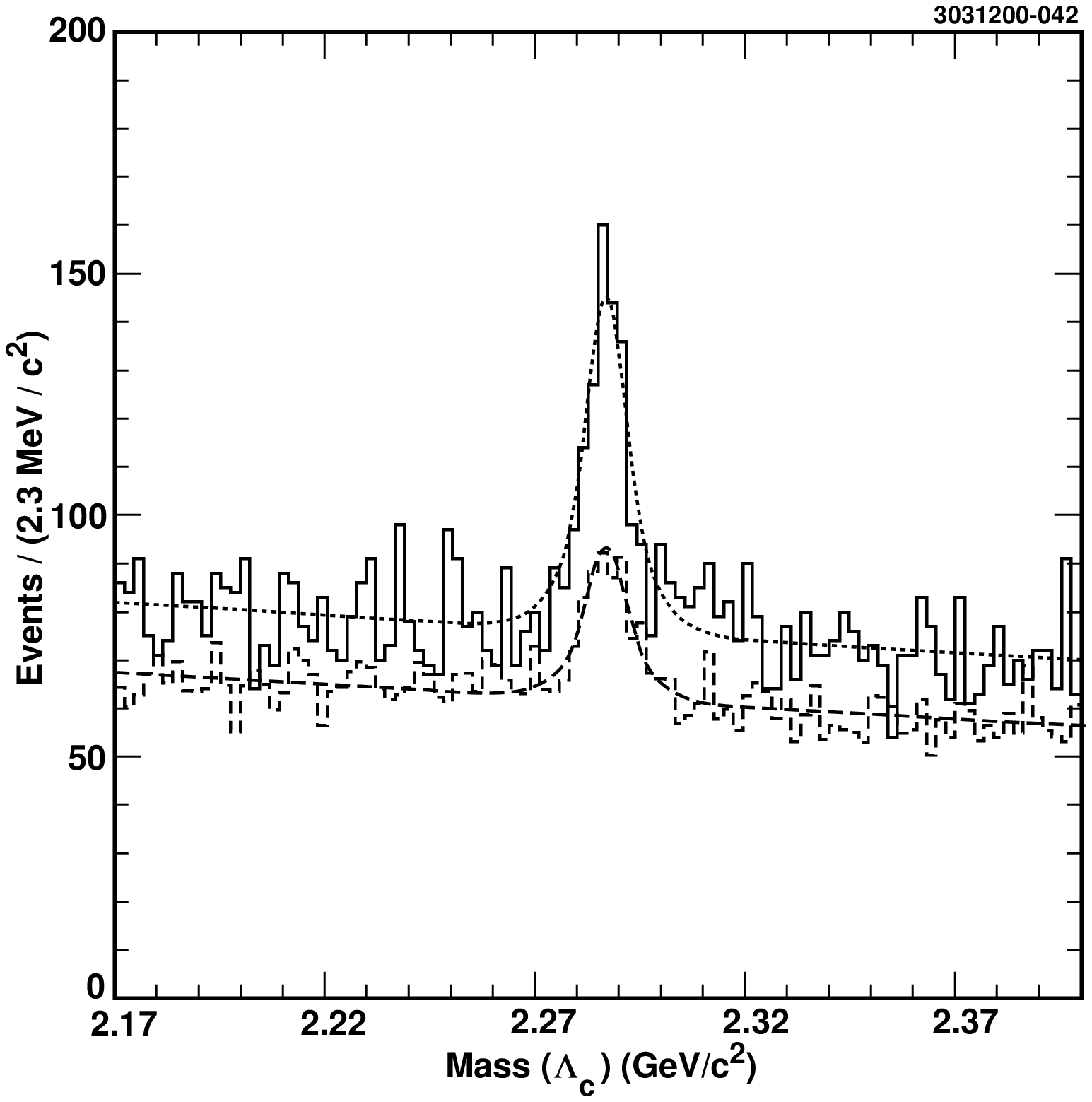}
\end{picture} \\
\caption{\small Projections of Figure 1 onto the candidate
$\Lambda_c^+$ mass, after requiring that the recoil mass be consistent
with the ${\overline\Lambda_c}^-$ signal (solid histogram,
with dotted curve fit overlaid) and sideband (dashed) regions
(see text for details). 
\label{fig:signal_sideband_overlay}}
\end{figure}

The total correlated double-tag yield is 
first determined by performing a 
sideband subtraction in the two-dimensional invariant mass
plot. 
Consider, for example Fig. \ref{fig:lclcb}.
We take one-dimensional projections of three slices in the
candidate ${\overline\Lambda_c}^-$ recoil
invariant mass ``$M_{recoil}$'' -
the ${\overline\Lambda_c}^-$ signal region:
($|M_{recoil}-2.286|<$0.03 $GeV/c^2$)
and the two ${\overline\Lambda_c}^-$ sideband regions:
($0.04<|M_{recoil}-2.286|<$0.07 $GeV/c^2$).
We then subtract the $\Lambda_c^+$
distributions from the ${\overline\Lambda_c}^-$ 
sidebands from that of the ${\overline\Lambda_c}^-$ signal region.
Figure \ref{fig:signal_sideband_overlay}. 
shows the signal distribution and the sum of the two sideband    
distributions.  
In performing these fits, the double-Gaussian signal shapes are 
constrained using the parameters determined from fits to the
single-tag sample.

We also perform a single fit in two dimensions to extract the signal yields.
In this latter fit, a two-dimensional Gaussian signal is used to parametrize
the peak region, two single Gaussians are used to fit the ``ridges'' away
from the peak region (corresponding to true charmed hadrons along one
axis in association
with combinatoric background on the other axis)
and a two-dimensional, smooth polynomial is used to
parametrize the background. 
Differences in i) the selection
of the sideband/signal regions and the regions projected, as
well as ii) the difference between the sideband-subtracted yield compared
to the two-dimensional, single-fit yield give a measure
of the fitting systematics for each
measurement.

\begin{table}[htpb]
\caption{\label{tab:Double-tags}Double Tag Yields obtained
from two-dimensional fits (statistical errors only).}
\begin{center}
\begin{small}
\begin{tabular}{lcccc}&&& Data & Monte Carlo\\
                                                         \hline
 \quad\quad$\Lambda_c^+ | {\overline\Lambda_c}^-$     &&& 
   253 $\pm$ 37 &  97 $\pm$ 40 \\
 \quad\quad$\Lambda_c^+ | {\overline D^0}$      &&& 
   722 $\pm$ 55 & 1905 $\pm$ 78 \\
 \quad\quad$\Lambda_c^+ | D^-$                 &&& 
   556 $\pm$ 71 & 1281 $\pm$ 96 \\
 \quad\quad$\Lambda_c^+ | {\overline\Sigma_c}$   &&& 
   34 $\pm$ 14 & 13 $\pm$ 9 \\
 \quad\quad$\Lambda_c^+ | {\overline \Lambda}$  &&& 
   1355 $\pm$ 72   & 1079 $\pm$ 85 \\             
\hline
 \quad\quad$\Sigma_c | {\overline D^0}$      &&& 
   29 $\pm$ 13 & 88 $\pm$ 13 \\
 \quad\quad$\Sigma_c | D^-$                 &&& 
   32 $\pm$ 18 & 59 $\pm$ 14 \\
 \quad\quad$\Sigma_c | {\overline \Lambda}$  &&& 
   122 $\pm$ 19   & 49 $\pm$ 12 \\             
\hline
 \quad\quad${\Lambda} | {\overline D^0}$              &&& 
   2400 $\pm$ 69 & 8547 $\pm$ 117 \\
 \quad\quad${\Lambda} | D^-$                         &&& 
   2132 $\pm$ 90 & 6397 $\pm$ 144 \\                            
\hline
\end{tabular}
\end{small}
\end{center}
\end{table}

We define the ``production rate'' (Table \ref{tab:Production-rates})
as the percentage of times 
we find one specific
particle in an event opposite a given tag (i.e., the
number of double tags, given
in Table \ref{tab:Double-tags} divided by the total number
of single tags, presented in Table \ref{tab:Single-tags}).
Production ratios are formed by taking ratios of 
production rates (Table \ref{tab:Ratios}). Note that, to compare
$\frac{{\Lambda}^{+}_{c} | 
{\overline\Lambda_c}^{-}}{{\overline\Lambda_c}^{-}}$ to
$\frac{{\Lambda}^{+}_{c} | \overline{D}}{\overline{D}}$, we must 
take into account the fact that the latter ratio implicitly
includes both sign combinations
($\frac{{\Lambda}^{+}_{c} | \overline{D}}{\overline{D}}$ plus
$\frac{{\overline\Lambda_c}^{-} | D}{D}$), while there is only one
unique way to form
$\frac{{\Lambda}^{+}_{c} | {\overline\Lambda_c}^-}{{\overline\Lambda_c}^-}$.
To compare these two rates and form production ratios
(Table \ref{tab:Ratios}),
the production rates for 
${\overline\Lambda_c}^-$-tags presented in the
table have therefore been multiplied by this factor of two.

\begin{table}[htpb]
\caption{\label{tab:Production-rates}Production Rates; statistical
errors only.}
\begin{center}
\begin{small}
\begin{tabular}{cccccc}                                                \hline
 \quad$\frac{Double \ tags}{Single \ tags}$ 
  && Data Fraction & Monte Carlo Fraction                \\ \hline
 \quad\quad$2\times\frac{\Lambda_c^+ | {\overline\Lambda_c}^-}{\overline\Lambda_c^-}$
   && (7.19 $\pm$ 1.08 ) x 10$^{-3}$ & (1.49 $\pm$ 0.62 ) x 10$^{-3}$ \\
 \quad\quad$\frac{\Lambda_c^+ | \overline{D^0}}{\overline{D^0}}$
   && (2.05 $\pm$ 0.16) x 10$^{-3}$ & (1.82 $\pm$ 0.08 ) x 10$^{-3}$ \\
 \quad\quad$\frac{\Lambda_c^+ | D^-}{D^-}$
   && (2.03 $\pm$ 0.26) x 10$^{-3}$ & (1.82 $\pm$ 0.14 ) x 10$^{-3}$ \\
 \quad\quad$\frac{\Lambda | {\overline\Lambda_c}^-}{\overline\Lambda_c^-}$ 
   && (19.3 $\pm$ 1.1) x 10$^{-3}$  & (8.28 $\pm$ 0.66) x 10$^{-3}$   \\ \hline
 \quad\quad$\frac{\Sigma_c | {\overline\Lambda_c}^-}{\overline\Lambda_c^-}$
   && (0.49 $\pm$ 0.20 ) x 10$^{-3}$ & (0.102 $\pm$ 0.066) x 10$^{-3}$ \\
 \quad\quad$\frac{\Sigma_c | \overline{D^0}}{\overline{D^0}}$
   && (0.082 $\pm$ 0.036 ) x 10$^{-3}$ & (0.068 $\pm$ 0.008) x 10$^{-3}$ \\
 \quad\quad$\frac{\Sigma_c | D^-}{D^-}$
   && (0.067 $\pm$ 0.049 ) x 10$^{-3}$ & (0.084 $\pm$ 0.020) x 10$^{-3}$ \\
 \quad\quad$\frac{\Lambda | {\overline\Sigma_c}}{\overline\Sigma_c}$ 
   && (32.1 $\pm$ 5.2) x 10$^{-3}$ & (7.6 $\pm$ 1.8) x 10$^{-3}$ \\
 \quad\quad$\frac{\Lambda | \overline{D^0}}{\overline{D^0}}$
   && (6.81 $\pm$ 0.20 ) x 10$^{-3}$ & (8.17 $\pm$ 0.11 ) x 10$^{-3}$ \\
 \quad\quad$\frac{\Lambda | D^-}{D^-}$
   && (7.79 $\pm$ 0.33 ) x 10$^{-3}$ & (9.07 $\pm$ 0.21 ) x 10$^{-3}$  \\
                                                                        \hline
\end{tabular}
\end{small}
\end{center}
\end{table}

\begin{table}[htpb]
\caption{Production Ratios}
\label{tab:Ratios}
\begin{center}
\begin{small}
\begin{tabular}{c|cc}
& Data (stat. and sys. errors) & Monte Carlo (stat. error only)\\                            \hline
 $\frac{2\times\Lambda_c^+ | 
{\overline\Lambda_c}^-}{\overline\Lambda_c^-} \div  
  \frac{\Lambda_c^+ | \overline{D^0}}{\overline{D^0}}$ 
   & {{(3.51 $\pm$ 0.59 $\pm$ 0.42)}} & {{(0.82 $\pm$ 0.34)}} \\
 $\frac{2\times\Lambda_c^+ | {\overline\Lambda_c}^-}{\overline\Lambda_c^-} 
\div 
  \frac{\Lambda_c^+ | {D^-}}{{D^-}}$
   & {{(3.54 $\pm$ 0.70 $\pm$ 0.43)}} & {{(0.80 $\pm$ 0.35)}} \\
 $\frac{{\Lambda} | {\overline\Lambda_c}^-}{\overline\Lambda_c^-} \div  
  \frac{{\Lambda} | \overline{D^0}}{\overline{D^0}}$
   & {{(2.83 $\pm$ 0.18 $\pm$ 0.26)}} & {{(1.01 $\pm$ 0.08)}} \\
 $\frac{{\Lambda} | {\overline\Lambda_c}^-}{\overline\Lambda_c^-} \div 
  \frac{{\Lambda} | {D^-}}{D^-}$
   & {{(2.48 $\pm$ 0.18 $\pm$ 0.23)}} & {{(0.91 $\pm$ 0.08)}} \\
 $\frac{{\Lambda} | {\overline\Sigma_c}}{\overline\Sigma_c} \div  
  \frac{{\Lambda} | \overline{D^0}}{\overline{D^0}}$
   & {{(4.71 $\pm$ 0.77 $\pm$ 0.56)}} & {{(0.93 $\pm$ 0.22)}} \\
 $\frac{{\Lambda} | {\overline\Sigma_c}}{\overline\Sigma_c} \div 
  \frac{{\Lambda} | {D^-}}{D^-}$
   & {{(4.12 $\pm$ 0.68 $\pm$ 0.52)}} & {{(0.84 $\pm$ 0.22)}} \\
                                                                        \hline
\end{tabular}
\end{small}
\end{center}
\end{table}

From Table \ref{tab:Production-rates}, we see a clear enhancement in 
the likelihood of producing a $\Lambda_c^+$ opposite a 
${\overline\Lambda_c}^-$
compared to a ${\overline D}$. This is observed in the case where
the $\Lambda_c^+$ is fully reconstructed, as well as tagged inclusively
by a $\Lambda$. Note that the fractional enhancement is smaller in 
the case when the $\Lambda_c^+$ is tagged inclusively. This is 
qualitatively consistent with the expectation that events containing two
charmed mesons and two baryons will produce a $\Lambda|{\overline D}$ 
correlation
which will inflate the denominator when we construct the ratio:
$\frac{{\Lambda} | {\overline\Lambda_c}^-}{{\overline\Lambda_c}^-}$ /
$\frac{{\Lambda} | \overline{D}}{\overline{D}}$. Non-cancelling
contributions to the numerator, such as 
$\Lambda K {\overline N} D | {\overline\Lambda_c}^-$, 
plus decays of charmed baryons other than $\Lambda_c$ into $\Lambda$,
may also be present.

\section{Cross Checks}
A number of cross checks were conducted in order to verify the accuracy of
the double tag signal extraction.
These include signal extractions of two-dimensional
``wrong sign'' plots (expected to have zero signal yield), 
as well as consistency checks with Monte Carlo and studies of
$D|{\overline D}$ correlations.

\subsection{Null and Wrong sign signals}
Neglecting the doubly Cabibbo suppressed
decay ${\overline D^0}\to K^-\pi^+$,
we expect a null signal yield from a 
double tag plot of ($m_{K^-\pi^+}|m_{K^-\pi^+}$). 
Similarly, we expect 
zero signal yield from a double tag plot of ($m_{K^-\pi^+}m_{K^+\pi^-}|$) in 
the case where the $D^0$ and ${\overline D^0}$ candidate 
are in the same hemisphere.
All of these
correlations give signal yields consistent with zero
as expected (Table 
\ref{tab:null-checks}).

\begin{table}[htpb]
\caption{\label{tab:null-checks}Null Checks (statistical errors only).}
\begin{tabular}{cc}
Correlation & Yield \\ \hline
 $D^0 | D^0$ (opposite hemisphere) & -16 $\pm$ 47 \\
$D^\pm|D^\pm$ (opposite hemisphere) & -10 $\pm$ 83 \\
 $D^0|D^+$ (opposite hemisphere) &  55 $\pm$ 91 \\
$D^0 {\overline D^0}|$ (same hemisphere) & -15 $\pm$ 25 \\
$D^\pm D^\mp|$ (same hemisphere) & -116 $\pm$ 157 \\ \hline
\end{tabular}
\end{table}

\subsection{Monte Carlo studies}
\begin{figure}[htpb]
\begin{picture}(200,250)
\includegraphics{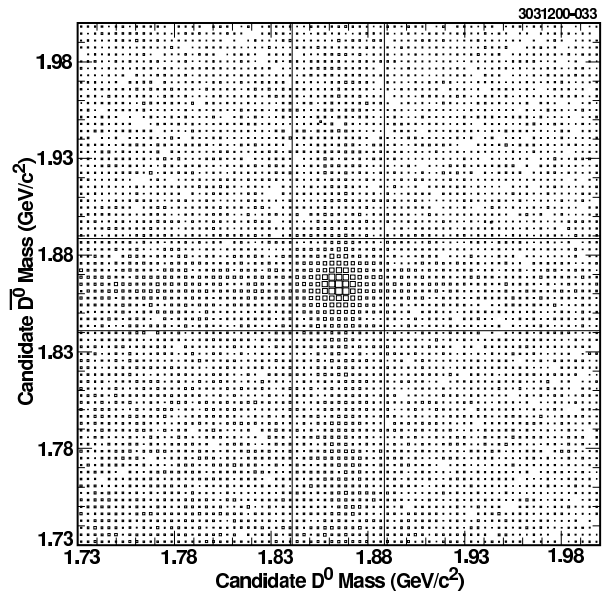}
\end{picture} \\
\caption{\small Double tag plot of $D^0 | \overline{D^0}$.
Horizontal and vertical solid lines designate signal bands.
}
\label{fig:d0d0b}
\end{figure}

The uncertainty in $\Lambda_c$ production characteristics is expected
to be somewhat larger than the uncertainty in $D^0$ production.
We therefore expect the Monte Carlo and data to agree on,
e.g., the number of $D^+ | D^-$ double tags per $D^-$ 
($\frac{D^+ | D^-}{D^-}$), 
and the number of $D^0 | \overline{D^0}$ double tags 
(Figure \ref{fig:d0d0b})
per ${\overline D^0}$
($\frac{D^0 | \overline{D^0}}{\overline D^0}$), as shown in Table 
\ref{tab:D-double-tags}.\footnote{The measured value of 0.0119, for instance,
is qualitatively consistent with the expectation that,
per ${\overline D^0}$ tag, 50\% of the time the charm quark will
produce a $D^0$, which is reconstructed in the $D^0\to K^-\pi^+$ mode
(${\cal B}(D^0\to K^-\pi^+)\sim$0.04) with approximately 60\% 
efficiency.}

Within errors, the agreement is good.
We have also further
ensured that there is no bias in the signal extraction
due to possible peaking of the background in the signal region, 
by subtracting all the true signal from a Monte Carlo double tag plot
and verifying that the measured yield, after subtraction of the true
generated particles, is indeed consistent with zero.

\begin{table}[htpb]
\caption{Consistency Check using $D|{\overline D}$ correlations
(statistical errors only).}
\label{tab:D-double-tags}
\begin{center}
\begin{small}
\begin{tabular}{c|c|c}
 \quad &\quad Data Raw Rate  & Monte Carlo Raw Rate    \\  \hline\hline
 \quad $D^0 | \overline{D^0}$ yield &  2099 $\pm$ 56 &  6238 $\pm$ 97\\
 \quad $\frac{D^0 | \overline{D^0}}{\overline{D^0}}$
production~rate &  0.0119 $\pm$ 0.0003 &  0.0119 $\pm$ 0.0002\\
                                                                        \hline
 \quad $D^+ | D^-$ yield &  1260 $\pm$ 88 & 3279 $\pm$ 130 \\
 \quad $\frac{D^+|D^-}{D^-}$ 
production rate &  0.0092 $\pm$ 0.0006 &  0.0093 $\pm$ 0.0004\\
                                                                        \hline
\end{tabular}
\end{small}
\end{center}
\end{table}

\subsection{Check of expected uncorrelated yield}
Having obtained values for production rates, we can 
compare these
with the
values expected under
the assumption of two hemispheres fragmentating independently. 
In this (hypothetical) limit, the ratio of double-tags to single-tags
should be the same as the ratio of single-tags to total charm quarks, since
each simply expresses the probability of a charm quark to evolve into
a particular final-state particle. 
Then,
using ${\cal B}(\Lambda_c\to X)$ to designate the branching fraction
for $\Lambda_c$ to decay into the final state $X$, 
${\cal L}$ as the integrated luminosity of our data set, and
$f_{(c \to \Lambda_c)}$ as the fraction of times a charm quark
materializes as a $\Lambda_c$ baryon, 
we can
relate the
expected number of single tags and double tags as:

       \[ \quad\quad\quad\quad\quad\quad
          N (\Lambda_c)_{Single~Tags} = {\cal L}\cdot\sigma(e^+e^-\to c\bar c)
          \cdot f_{(c \to \Lambda_c)} \cdot {\cal B}(\Lambda_c \to X) 
          \cdot {\epsilon}_{\Lambda_c\to X}
       \quad\quad\quad\quad\quad (2) \]
and, for the probability that, given a reconstructed ${\Lambda_c}$,
we reconstruct an opposite ${\overline\Lambda_c}^-$
       \[ \quad\quad\quad
          N ({\Lambda_c^+ | {\overline\Lambda_c}^-)_{Double~Tags} = 
          f^{'}_{(c \to \Lambda_c)} \cdot {\cal B}(\Lambda_c \to X) 
          \cdot {\epsilon}^{'}_{\Lambda_c\to X} 
\cdot N(\Lambda_c)_{Single~Tags}}
       \quad\quad\quad  (3)\]\\

In equation (2), for the number of single $\Lambda_c^+$
tags, we use half the
total number of observed $\Lambda_c^++{\overline\Lambda_c}^-$ (70199/2).
In equation (3), the 
term $f^{'}_{(c \to \Lambda_c)}$ indicates that the 
fraction of times that a charm quark produces a $\Lambda_c$ 
may be different if the event already 
contains a ${\overline\Lambda_c}$ produced 
from the corresponding anticharm quark. This is, of course,
exactly the correlation
factor we wish to ultimately measure.  
The prime on ${\epsilon}^{'}_{\Lambda_c\to X}$ indicates 
that the efficiency for reconstructing a 
$\Lambda_c$ may be higher for events in 
which a ${\overline\Lambda_c}$ has 
already been reconstructed due to geometrical correlations --
since the charm and anticharm quarks are back 
to back in an $e^+e^- \to$ c\=c event, 
reconstruction of one charmed particle 
ensures that the
corresponding antiparticle is in a good acceptance region of the detector.
Dividing equation (3) by equation (2) and 
multiplying each side by $N(\Lambda_c)_{Single~Tags}$ 
we obtain, for the number of double tags:
       \[ \quad\quad\quad\quad\quad\quad
          N ({\Lambda_c^+ | {\overline\Lambda_c}^-})_{Double Tags} = 
          \frac{(N (\Lambda_c)_{Single Tags})^{2} \cdot f^{'}_{(c \to 
          \Lambda_c)} \cdot {\epsilon}^{'}_{\Lambda_c\to X}}{
{\cal L}\cdot\sigma(e^+e^-\to c\bar c)
          \cdot f_{(c \to \Lambda_c)} \cdot {\epsilon}_{\Lambda_c\to X}}
      \quad\quad\quad\quad\quad\quad  (4)\]\\
The difference between $f_{(c \to \Lambda_c)}$ 
and $f^{'}_{(c \to \Lambda_c)}$ 
represents the enhancement in production 
of the $\Lambda_c$ when a 
$\overline{\Lambda_c}$ is present.  
With
$\frac{{\epsilon}^{'}_{\Lambda_c\to X}}{{\epsilon}_{\Lambda_c\to X}}
\equiv{\epsilon}_{geometry}$
and $\frac{f^{'}_{(c \to \Lambda_c)}}{f_{(c \to \Lambda_c)}}
\equiv f_{correlated}$
(=1 if independent fragmentation holds), we 
can estimate our
expected double tag to single tag ratio as follows:
From the luminosity, we calculate the number of 
$e^+e^-\to c{\overline c}$ events using as inputs
the $e^+e^-\to q{\overline q}$ cross-section (3.3 nb), and taking
$\frac{e^+e^-\to c{\overline c}}{e^+e^-\to q{\overline q}}$=0.4. Assuming
that $f_{correlated}$ is equal to unity for $D|{\overline D}$
events, we can solve Eqn. (4) for
${\epsilon}_{geometry}$ by comparing the 
total number of $D^0 | \overline{D^0}$ and $D^+ | D^-$ double tags 
we would expect to find (assuming independent fragmentation) with the
actual number of $D^0 | \overline{D^0}$ and $D^+ | D^-$ double tags.
This is calculated to be 
$${\epsilon}_{geometry}={(2099)(0.4)(3.3)(13.6\times 10^6)
\over (352294/2)^2}\equiv
1.21\pm0.04~(statistical~errors~only),$$ using
the total luminosity ${\cal L}$=13.6 fb$^{-1}$,
and the single tag and double tag values for our $D^0$ and 
$D^0|{\overline D^0}$ samples, respectively. For the
$D^+$ and $D^+|D^-$ samples, we obtain ${\epsilon}_{geometry}=1.21\pm0.09$.
Using a value of ${\epsilon}_{geometry}\equiv 1.21$, 
a similar calculation can be performed for the expected number of
${\Lambda_c^+ | {\overline\Lambda_c}^-}$ double tags, 
based on the total number of c\=c events and the single
tag yield, giving
an expected value of 83.0$\pm$2.6 (statistical
error only) double tag events.
This number is then compared to the 
measured number of ${\Lambda_c^+ | {\overline\Lambda_c}^-}$ 
double tags to obtain an estimate of $f_{correlated}={253\pm37\over
83.0\pm2.6}=3.04\pm0.45$.
Despite the roughness of this
approach, the observed enhancement is consistent with our
calculated production ratios.
%

\section{Study of the $\Lambda_c^+ | {\overline\Lambda_c}^-$ signal
characteristics}
\begin{figure}[htpb]
\begin{picture}(200,250)
\includegraphics{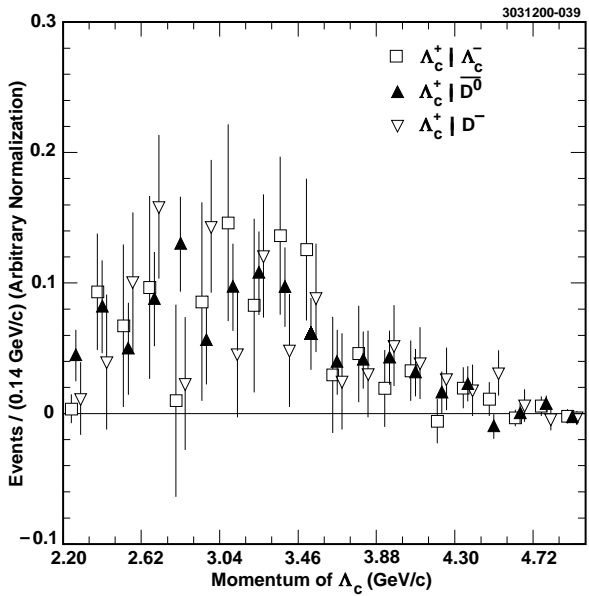}
\end{picture} \\
\caption{\small $\Lambda_c$ (and charge conjugate)
momentum distribution for the 
three double tag signals 
$\Lambda_c^+ | {\overline\Lambda_c}^-$, 
$\Lambda_c^+ | \overline{D^0}$, and $\Lambda_c^+ | D^-$.  
Squares represent the $\Lambda_c^+ | {\overline\Lambda_c}^-$ signal, 
upright triangles represent the $\Lambda_c^+ | \overline{D^0}$ signal, 
and inverted triangles represent the $\Lambda_c^+ | D^-$ signal.  
These three distributions are qualitatively similar, 
indicating similar production dynamics. 
All distributions have been normalized to have unit area.
\label{fig:dtag-mom-correl}}
\end{figure}

\begin{figure}[htpb]
\begin{picture}(200,250)
\includegraphics{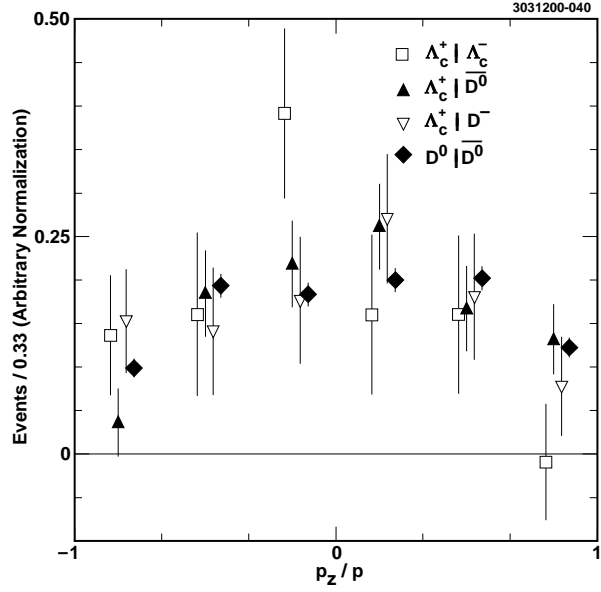}
\end{picture} \\
\caption{\small Polar angle distribution 
with respect to the $e^+e^-$ beam axis ($p_z$/$p$) 
for the
double tag signals $\Lambda_c^+ | {\overline\Lambda_c}^-$ (squares), 
$\Lambda_c^+ | \overline{D^0}$ (triangles), $\Lambda_c^+ | D^-$
(inverted triangles), and
$D^0|{\overline D^0}$ (diamonds).
All distributions have been normalized to have unit area.}
\label{fig:dtag-polar-correlA}
\end{figure}

\begin{figure}[htpb]
\begin{picture}(200,250)
\includegraphics{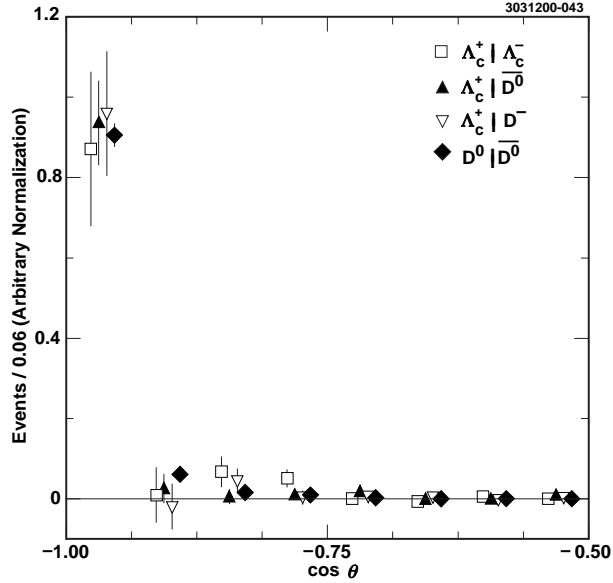}
\end{picture} \\
\caption{\small Production angle between 
the $\Lambda_c^+$ and ${\overline\Lambda_c}^-$ (squares) compared
with $\Lambda_c^+|{\overline D^0}$ (upright triangle),
$\Lambda_c^+|D^-$ (inverted triangle), and
$D^0|{\overline D^0}$ (diamond) event
samples. As expected,
$\Lambda_c^+$ are produced back-to-back with respect to
${\overline \Lambda_c}^-$ tags.
All distributions have been normalized to have unit area.}
\label{fig:dtag-openingangle-correl}
\end{figure}

\begin{figure}
\begin{picture}(200,250)
\includegraphics{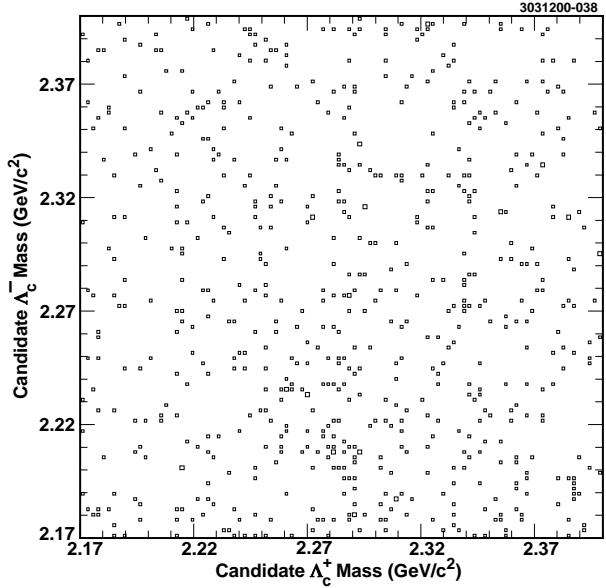}
\end{picture} \\
\caption{\small Double-tag $\Lambda_c^+ | 
{\overline\Lambda_c}^-$ signal in events containing an identified
antiproton or proton.}
\label{fig:fourbaryon}
\end{figure}

We have compared some of the production characteristics of the
observed $\Lambda_c^+ | {\overline\Lambda_c}^-$ signal with the
$\Lambda_c^+ | \overline{D^0}$ and $\Lambda_c^+ | D^-$ signals.
No obvious difference is found in either 
the $\Lambda_c^+$ momentum 
spectrum (or ${\overline\Lambda_c}^-$, in
the charge conjugate case)
or the
polar angle distribution of the 
$\Lambda_c^+$ 
(see Figs. \ref{fig:dtag-mom-correl} and
\ref{fig:dtag-polar-correlA}, respectively)
for the 
${\overline\Lambda_c}^-$-tagged or 
the ${\overline D}$-tagged samples.
We also find that the observed
$\Lambda_c^+ | {\overline\Lambda_c}^-$ cross-section is statistically
the same for the
data taken on the $\Upsilon$(4S) resonance as data taken just below the
$\Upsilon$(4S). As expected (Figure 
\ref{fig:dtag-openingangle-correl}) if the
charmed baryons are following the direction of the original
charm/anticharm quarks, the observed opening angle between the
$\Lambda_c^+$ and ${\overline\Lambda_c}^-$ peaks at 180$^\circ$.
Finally,
to investigate the possibility that the $\Lambda_c^+ | {\overline\Lambda_c}^-$
signal was associated with the production of a 4-baryon system, we 
plot the double tag yield in cases where there are well-identified protons
found in the same event (Fig. 
\ref{fig:fourbaryon}). No signal is observed in such a case.

\section{$\Sigma_c$ $({1\over 2})^+$ production}
\begin{figure}[htpb]
\begin{picture}(200,250)
\includegraphics{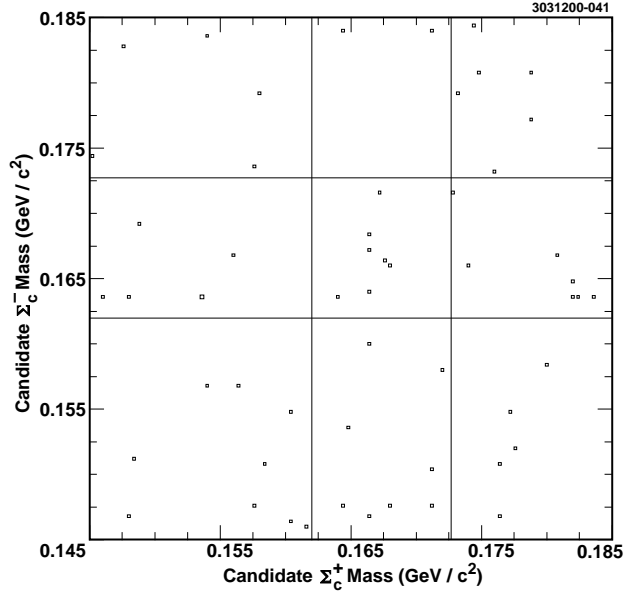}\
\end{picture} \\
\caption{\small Double-tag $\Sigma_c | 
{\overline\Sigma_c}$ mass difference plot.
Horizontal and vertical solid lines designate signal bands.
}
\label{fig:ScScb.ps}
\end{figure}

We have also investigated the possibility of correlated
$\Sigma_c | {\overline\Sigma}_c$ production. Of particular interest here
is the possibility of observing an enhanced production rate
relative to our observed $\Lambda_c|{\overline\Lambda_c}^-$ correlation.
Such an enhancement may be indicative of spin correlations at the
first rank in fragmentation.
$\Sigma_c$'s
are reconstructed through
their decay mode into $\Lambda_c$ and a soft pion
$\pi_s$: $\Sigma_c\to\Lambda_c\pi_s$. For the highest efficiency
and best resolution, only charged pions $\pi^\pm_s$ are used in reconstructing
$\Sigma_c$'s. 
Although statistically limited, a direct
search for $\Sigma_c|{\overline\Sigma_c}$ double tags yields 6 events in the
signal region (Figure \ref{fig:ScScb.ps}); 
the extrapolated background under the signal is
$1.8\pm0.3$ events. 

Using the total number of $\Sigma_c$ single tags
observed ($3804\pm185$), the total number of $\Lambda_c$ single
tags observed ($70199\pm1604$), and the total number of
$\Lambda_c|{\overline\Lambda_c}^-$ double tags observed ($252.5\pm37.4$), 
we can estimate the number of expected 
$\Sigma_c|{\overline\Sigma_c}$ double tags 
(assuming that the $\Sigma_c|{\overline\Sigma_c}$ correlated production
rate is the same as the correlated $\Lambda_c|{\overline\Lambda_c}^-$ 
production
rate) as: 
252.5$\times ({3804\over 70199})^2$=$0.74\pm 0.11$. We therefore see a 
statistically very limited indication of $\Sigma_c|{\overline\Sigma_c}$ 
correlations, suggesting a spin correlation at the first step of fragmentation.
More data are necessary to elucidate this situation.

\section{Systematics}
We expect most of the systematic uncertainties associated with 
measurement of both numerator and
denominator of the production
rates to cancel in calculation of the production ratios.
The sensitivity to the definition of ``signal'' and ``sideband'' 
mass regions
in the one- and two-dimensional subtraction was determined by calculating
the yields as we vary these parameters. 
We assign a systematic error of 6\% due to ``signal
parameterization uncertainty'' based on this study. 
We have also
performed a second (complete) set of fits using two-dimensional 
Gaussian parameterizations of the signal and two-dimensional polynomial
parameterizations of the background. The two techniques show excellent 
agreement, and are typically within 5\% of each other, for each
two-dimensional extraction.
When we vary the selection criteria 
(e.g., particle identification criteria, and minimum momentum requirements)
used to define our hadrons ($\Lambda_c$, $D^0$, and $D^+$), we observe
a maximum 7\% variation in the production ratios we calculate.
As an example of our investigation of the momentum dependence of the 
$\Lambda_c^+{\overline\Lambda_c}^-$
production ratio enhancement, Table \ref{tab:mom_correl} shows the
production ratio in the two-dimensional $(p_{\Lambda_c},p_{\overline H_c})$
space. 

\begin{table}[htpb]
\caption{\label{tab:mom_correl}Production ratio enhancement tabulated as
a function of $\Lambda_c$ momentum
and also the tag momentum in the production rate
denominator. For example,
the upper left entry ($3.02\pm0.95$) represents the
production ratio 
 $\frac{\Lambda_c^+ | {\overline\Lambda_c}^-}{\overline\Lambda_c^-} \div  
  \frac{\Lambda_c^+ | \overline{D^0}}{\overline{D^0}}$, for the case where
the $\Lambda_c$, ${\overline\Lambda_c}^-$, and ${\overline D^0}$ each
have momentum between 2.3 and 3.3 GeV/c. Statistical
errors only are shown.}
\begin{tabular}{c|c|c} \\
 & 2.3 GeV/c$<p(\Lambda_c)<$3.3 GeV/c & 3.3$<p(\Lambda_c)<$5.0 GeV/c \\ \hline
2.3 GeV/c$<p({\overline\Lambda_c}^-)$, $p({\overline D^0})<3.3$ GeV/c &
 $3.02\pm0.95$ & $4.04\pm1.93$ \\
3.3 GeV/c$<p({\overline\Lambda_c}^-)$, $p({\overline D^0})<$5.0 GeV/c &
$2.60\pm1.23$ & $3.79\pm1.51$ \\ \hline
2.3 GeV/c$<p({\overline\Lambda_c}^-)$, $p(D^-)<$3.3 GeV/c & 
$3.30\pm1.33$ & $3.42\pm1.99$ \\
3.3 GeV/c$<p({\overline\Lambda_c}^-)$, $p(D^-)<$5.0 GeV/c 
& $3.39\pm1.76$ & $3.50\pm1.47$ \\ \hline
\end{tabular}
\end{table}
The systematic errors are added, mode-by-mode, to determine the overall 
systematic error for each calculated production ratio (Table \ref{tab:Ratios}).

\section{Summary}
The measured values of the production ratios imply that a $\Lambda_c$ is
roughly three times more likely to be produced opposite a
$\overline{\Lambda_c}$ than opposite either a ${\overline D^0}$ or a 
$D^-$. These
results indicate strong evidence in support of correlated production of the
$\Lambda_c$. Assuming that 6\% of charm quark fragmentation at
$\sqrt{s}\sim$10 GeV results in
a charmed baryon, our result implies that approximately 20\% of all
$\Lambda_c$'s are produced in association with a
${\overline\Lambda_c}^-$ in the opposite hemisphere. 
The effect is not produced by the Jetset simulation but does
not contradict the Lund String Model.  In fact if the primary $c\overline c$
were broken at very small proper time by diquark-antidiquark tunneling from
the vacuum, producing a (possibly excited) charm baryon system and anti-charm
antibaryon system, a correlation of the type observed here would be generated.

\section{Acknowledgments}
We gratefully acknowledge the effort of the CESR staff in providing us with
excellent luminosity and running conditions.
M. Selen thanks the PFF program of the NSF and the Research Corporation, 
A.H. Mahmood thanks the Texas Advanced Research Program,
F. Blanc thanks the Swiss National Science Foundation, 
and E. von Toerne thanks the Alexander von Humboldt Stiftung for support.
This work was supported by the National Science Foundation, the
U.S. Department of Energy, and the Natural Sciences and Engineering Research 
Council of Canada.

\end{document}